\documentclass{aa}
\usepackage{natbib}
\usepackage{txfonts}
\usepackage{graphicx}

\newcommand{\gflog}{\ensuremath{\log gf}}
\newcommand{\feh}{\ensuremath{\protect\rm [Fe/H] }}
\newcommand{\teff}{T$_{\rm eff}$}
\newcommand{\glog}{log\,g}
\newcommand{\kms}{\,$\mathrm{km\,s^{-1}}$}
\newcommand{\xx}{\ensuremath{\mathrm{1D}_{\mathrm{LHD}}}}
\newcommand{\mD}{\ensuremath{\left\langle\mathrm{3D}\right\rangle}}
\newcommand{\mlp}{\ensuremath{\alpha_{\mathrm{MLT}}}}

\newcommand{\cobold}{{\sf CO$^5$BOLD}}
\newcommand{\linfor}{{\sf Linfor3D}}
\newcommand{\Nt}{\ensuremath{N_\mathrm{t}}}
\newcommand{\tchar}{\ensuremath{t_\mathrm{c}}}

\idline{91}{1003}

\begin{document}
\title{\ion{Cu}{i} resonance lines 
in turn-off stars of NGC 6752 and NGC 6397.\thanks{Based on observations made with the 
ESO Very Large Telescope
at Paranal Observatory, Chile 
(Programmes 71.D-0155, 75.D-0807, 76.B-0133)}}
\subtitle{Effects of granulation from CO5BOLD models.}

\author{P. Bonifacio\inst{1,2,3}
\and
E. Caffau\inst{2,4}\thanks{Gliese Fellow}
\and
H.-G. Ludwig  \inst{1,2,4}
}
\institute{
CIFIST Marie Curie Excellence Team
\and
GEPI, Observatoire de Paris, CNRS, Universit\'e Paris Diderot; Place
Jules Janssen 92190
Meudon, France
\and
Istituto Nazionale di Astrofisica,
Osservatorio Astronomico di Trieste,  Via Tiepolo 11,
I-34143 Trieste, Italy
\and
Zentrum f\"ur Astronomie der Universit\"at Heidelberg, Landessternwarte, K\"onigstuhl 12, 69117 Heidelberg, Germany
}
\authorrunning{Bonifacio et al.}
	\titlerunning{Copper resonance lines}
\offprints{Piercarlo.Bonifacio@obspm.fr}
\date{Received 21 July  2009; Accepted 30 August 2010}

\abstract
{Copper is an element whose interesting evolution
with metallicity is not fully understood. Observations
of copper abundances rely on a very limited number of lines,
the strongest are the \ion{Cu}{i} 
lines of Mult.\,1 at 324.7\,nm and 327.3\,nm 
which can be measured even at extremely low metallicities.}
{We   investigate the quality of these  lines
as abundance indicators.}
{We measure these lines in two  turn-off (TO) stars 
in the Globular Cluster NGC 6752
and two TO stars in the Globular Cluster NGC 6397 and derive abundances
with 3D hydrodynamical  model atmospheres computed with the
{\sf CO5BOLD} code.
These abundances are compared to the Cu abundances measured
in giant stars of the same clusters, using the lines of Mult.\,2 at 510.5\,nm
and 578.2\,nm.}
{The abundances derived from the lines of Mult.\,1 \relax in TO stars
differ from the abundances of giants of the same clusters.
This is true both using {\sf CO5BOLD} models and using traditional
1D model atmospheres. The LTE 3D corrections for TO stars
are large, while they are small for giant stars.}
{The \ion{Cu}{i} resonance lines of  Mult.\,1 are not reliable
abundance indicators.  It is likely that departures
from LTE should be taken into account to properly describe
these lines, although it is not clear if these
alone can account for the observations. 
An investigation of these departures is indeed 
encouraged for both dwarfs and giants. Our recommendation
to those interested in the study of the evolution of copper
abundances is to rely on the measurements in giants, based on
the lines of Mult.\,2. We caution, however,
that NLTE studies may imply a revision in  all the
Cu abundances, both in dwarfs and giants.}

\maketitle 


\section{Introduction}

There is no wide consensus on the nucleosynthetic
origin of copper, and 
the complex picture drawn by the observations
has no straightforward interpretation.
Multiple channels can contribute to
the production of this element. According
to \citet{bisterzo} there are five such channels:
explosive nucleosynthesis, either in 
Type II supernovae (SNII) or in Type Ia 
supernovae (SNIa), slow neutron capture ($s-$process), either
weak (i.e.  taking place in massive
stars in conditions of hydrostatic equilibrium
during He and C burning) or main (i.e. 
occurring in the inter-shell region of 
low-mass asymptotic giant branch stars)
and the  {\it weak} $sr$-process.
The latter occurs in massive stars in the C-burning
shell when neutron densities reach very high
values, intermediate between typical 
$s-$process neutron densities ($10^9 - 10^{11}$
cm$^{-3}$; \citealt{despain}) and
$r-$process neutron densities ($10^{20} - 10^{30}$
cm$^{-3}$; \citealt{kratz}).
The contribution of the $s-$process, both weak and main,
to the solar system Cu abundance is estimated by
\citet{travaglio} to be 27\%.
Explosive nucleosynthesis in SNII can account
for 5\% to 10\% of the solar system Cu \citep{bisterzo}.
The contribution of SNIa is probably  less
well known, however, as pointed out
by \citet{McWilliam}, the available SNIa yields of Cu
 are rather low \citep{travagliosn,Thielemann}. 
\citet{bisterzo} claim that the bulk
of cosmic Cu has indeed been produced by the {\it weak} 
$sr$-process.

Observations of copper abundances
in Galactic stars show a decrease in
the Cu/Fe ratio at low metallicities.
This was first suggested
by \citet{Cohen80}, on the
basis of the measurements in giant
stars of several Globular Clusters 
of different metallicities
\citep{Cohen78,Cohen79,Cohen80}.
It was not until the comprehensive study 
of \citet{Sneden91} that this trend
was clearly defined in a robust way,
resting on measurements in a large sample of stars.
Recent studies of field  \citep{Mishenina,Bihain}
and  Globular Cluster stars 
\citep{Shetrone01,Shetrone03,Simmerer,Yong}
have confirmed this
trend \citep[see Fig.\,1 of][]{bisterzo}.
Somewhat at odds with these general results
are the observations of the Globular Cluster
$\omega$ Cen \citep{Cunha,Pancino}.
Even though this cluster 
shows a sizeable spread in metal
abundances \citep[$-2.20\le \feh \le-0.70$,][]{Johnson},
the Cu/Fe abundance ratio is nearly constant, with no
discernible trend.

Observations in Local Group galaxies
\citep{Shetrone01,Shetrone03} show that metal-poor
populations display low Cu/Fe ratios, similar
to what is observed in Galactic stars of comparable
metallicity. However, \citet{McWilliam} noted
that the metal-rich population of the Sgr dSph
displays considerably {\em lower} Cu/Fe
ratios than Galactic stars of comparable
metallicity. This  result is  confirmed by 
the measurements of \citet{Sbordone}, who also include
stars of the Globular Cluster Terzan 7, associated
to the Sgr dSph.

\begin{figure}
\resizebox{\hsize}{!}{\includegraphics[clip=true,angle=0]{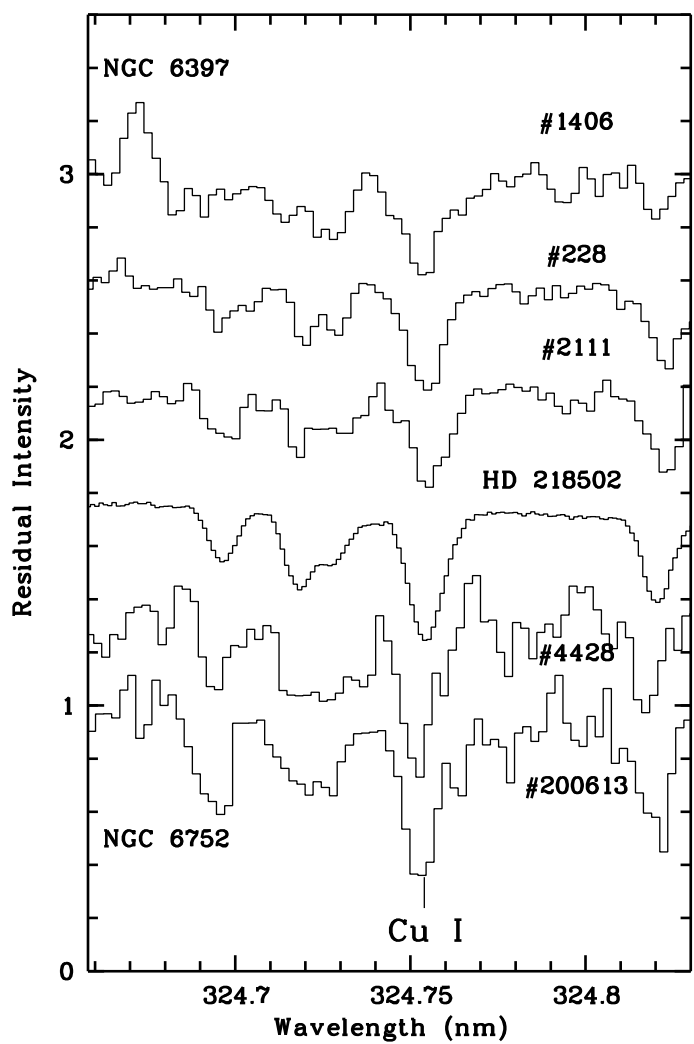}}
\caption{ \ion{Cu}{i}
324.7\,nm  line in the programme stars. The spectra
are displaced vertically by 0.4 units, with respect to each
other, for display purposes.\label{obs3247}}
\end{figure}

\begin{figure}
\resizebox{\hsize}{!}{\includegraphics[clip=true,angle=0]{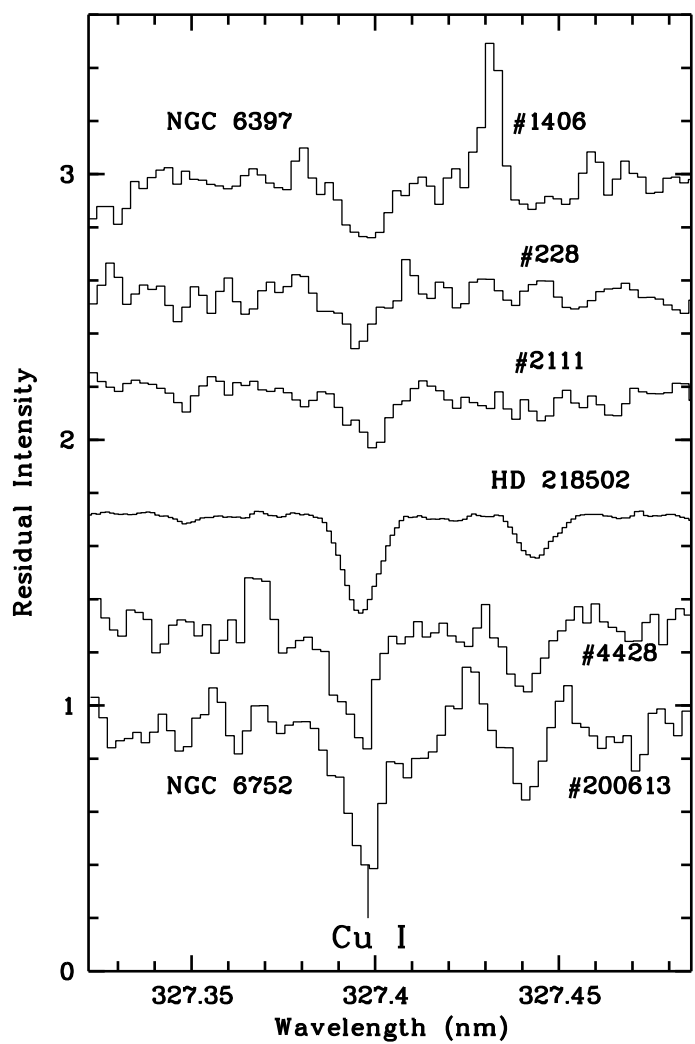}}
\caption{ \ion{Cu}{i}
327.3\,nm  line in the programme stars. The spectra
are displaced vertically by 0.4 units, with respect to each
other, for display purposes.\label{obs3273}}
\end{figure}

The majority of the Cu measurements
are based on the \ion{Cu}{i} lines of 
Mult.\,2\footnote{we refer to the multiplet designation of \citet{moore}},
sometimes one line of Mult.\,7 is used.
The  exceptions are  the measurements  of \citet{Bihain} and
\citet{Cohen08},
who use the resonance lines of Mult.\,1 and
\citet{prochaska} who, to our knowledge, are the only
ones who have made use of the strongest
line of Mult.\,6 in the near infrared.
While for stars of metallicity above --1.0 one may
have a choice of several
lines to use, when going to metal-poor stars, 
for instance below --1.5, the only usable Cu abundance
indicators are the two strongest lines
of \ion{Cu}{i} Mult.\,2 in giant stars
and the resonance lines of Mult.\,1 in 
both  dwarfs and giants.
The advantage of the lines of Mult.\,1 is that
they are very strong, at high metallicity they
are strongly saturated and therefore not ideal
for abundance work, but,
they remain measurable down to an extremely low
metallicity. \citet{Bihain} have been able to measure
the 327.3\,nm line in the extremely metal-poor dwarf 
G64-12 ([Fe/H]$\sim -3$).
Observationally the main disadvantage of Mult.\,1
is that it lies in the UV, fairly near to the atmospheric
cut-off. 
A very efficient UV spectrograph, like UVES and a 
large telescope, like VLT, may circumvent this problem.
There are many spectra suitable for the measurement
of the \ion{Cu}{i} lines of Mult.\,1 \relax in stars 
of different metallicities  in the ESO
archive\footnote{http://archive.eso.org}.
The main purpose of our investigation is to assess the
quality of the \ion{Cu}{i} lines of Mult.\,1 
as abundance indicators.
Our strategy is to compare for the first time Cu
abundances in
main sequence and giants of the same cluster, because  Cu is 
not expected to be
easily destroyed or
created. 
This test will indicate the reliability of
our modelling.
 Globular Clusters
NGC 6397 and NGC 6752 span an interesting range
in metallicity --2.0 to --1.5, which is relevant for a large
fraction of the observations in field stars.

\begin{table}
\caption{Atmospheric parameters of the programme stars.\label{tabatm}}
\begin{tabular}{lcccc}
\hline\hline\noalign{\smallskip}
Star  & T$_{\rm eff}$ & $\log g$   & [Fe/H] & $\xi$\\
      &     K         & [cgs]      &  dex   & \kms \\
\hline\noalign{\smallskip}
Cl* NGC 6752 GVS  4428     & 6226      &  4.28  & -1.52 & 0.70 \\
Cl* NGC 6752 GVS 200613    & 6226      &  4.28  & -1.56 & 0.70 \\
Cl* NGC 6397 ALA 1406      & 6345      &  4.10  & -2.05 & 1.32 \\
Cl* NGC 6397 ALA 228       & 6274      &  4.10  & -2.05 & 1.32 \\
Cl* NGC 6397 ALA 2111      & 6207      &  4.10  & -2.01 & 1.32 \\
HD 218502          & 6296      &  4.13  & -1.85 & 1.00 \\
\hline
\end{tabular}
\end{table}

\section{Observations and equivalent width measurements}

The spectra analysed here are the same as 
in \citet{Pasquini04} and \citet{Pasquini07}
and described in the above papers.
They were obtained with the UVES spectrograph \citep{dekker}
at the ESO VLT-Kueyen 8.2m telescope.
We  here use the blue arm spectra, which
are centred at 346\,nm. Both clusters
were observed with a $1''$ slit and a $2\times 2$
on-chip binning, which yields a resolution of about 40\,000.
The reduced spectra were downloaded from the ESO archive, 
thanks to the improved strategies for optimal
extraction \citep{ballester}, the S/N ratios
are greatly improved compared to 
what was previously available.
The equivalent widths (EWs) of the two \ion{Cu}{i} lines
of Mult.\,1 were measured with the IRAF task {\tt splot}
and are provided in the on-line Table \ref{labun}.

In addition to the four cluster stars we analyse
the field star HD 218502 as a reference. 
Its  atmospheric parameters are 
close to those of the cluster stars.
For this star we  work with the data used by
\citet{Pasquini04} as well as with the data
observed in 2005, in the course of ESO programme
76.B-0133 \citep[see][]{rodolfo}.
For this star we used six spectra:
two with  1\farcs{0} and $2\times 2$ binning, 
two with  1\farcs{0} and $1\times 1$ binning, 
two with  1\farcs{2} and $1\times 1$ binning. 
Each pair of spectra was coadded, the equivalent
widths were measured on the coadded spectrum and 
then the  three equivalent widths were averaged.
The spectra of all the five stars analysed here
are shown in Fig.\,\ref{obs3247} and \ref{obs3273}.

One of the goals of the present analysis is to compare
the Cu abundances in the TO stars with those measured
in giant stars of the same cluster.
For NGC 6752 we can rely on the recent analysis
by \citet{Yong}, who analysed 38 giants in this
cluster, making use of high-resolution high-S/N ratio
UVES spectra.
The atomic data used by \citet{Yong} are the same
as those here used. 
The analysis is based on 1D ATLAS models and LTE
spectrum synthesis. That the ATLAS models
employed
 by \citet{Yong} use the approximate overshooting
option in ATLAS, while those we use do not, brings
about only minor differences. 
Thus the measurements of  \citet{Yong}
are directly comparable to our own.
The measurements of \citet{Yong} are, however, based
only on the strongest line of Mult.\,2.
Because we aim to compare the abundance
derived from the lines of Mult.\,1 and Mult.\,2
we retrieved UVES reduced spectra of one of the
stars of \citet{Yong} from the ESO archive: Cl* NGC 6752 YGN 30.
We used  three spectra of 1800\,s
obtained with the dichroic \# 1,
the blue arm spectrum was centred at
346\,nm  and the red arm spectrum at 580\,nm.
The slit was set at 1\farcs{0} in the blue arm
and 0\farcs{7} in the red arm; the
CCD binning was $1\times1$ for both arms.
The corresponding resolution is $\sim 45\,000$
for the blue arm and $\sim 60\,000$ in the red arm.

For the cluster NGC 6397, though, we were unable to 
find any recent analysis that included the measurement
of Cu. In fact the only measurement of Cu in this
cluster which we could find is due to \citet{gratton82}.
In order to make the \citet{gratton82} measurements 
directly comparable to our own we used the published
EWs of the \ion{Cu}{i} lines of Mult.\,2 and derive
the abundances with our models, spectrum synthesis
codes, and atomic data.

\section{Cu abundances}

\subsection{Atomic data}

To determine the Cu abundances for the TO stars
we used 
the \ion{Cu}{i} resonance lines of Mult.\,1 
at 324.7\, nm  and 327.3\, nm.
The \gflog\ values were taken from 
\citet{bielski} and
the hyperfine structure and isotopic shifts
for the $^{63}$Cu and $^{65}$Cu isotopes 
from \citet{kuruczsite}.
We used the same sources for the two lines
of Mult.\,2 at 510.5\, nm and 587.2\,nm 
that we used for the giant stars.
The line list used for the computations is given
on-line in   Table \ref{tabgf}.
The line at  327.3\, nm is free from
blends in metal-poor TO stars and 
the continuum is usually  easily determined.
The stronger 324.7\, nm line, though,
lies in a more complex spectral region.
The only truly blending feature is a weak 
OH line (324.7615\,nm), but, the line
is on the  red wing of a complex blend, mainly
of iron lines, of which several have poor \gflog\
values. The continuum is more difficult to determine
for this line, given the larger line crowding in this region.
We experimented with different choices for the 
Van der Waals broadening of the lines, the
ABO theory \citep{abo1,abo2,abo3,abo4} and the
WIDTH approximation 
(\citealt{1993KurCD..13.....K,2005MSAIS...8...14K,2005MSAIS...8...44C}, 
see also \citealt{ryan}).
For the transitions under consideration the 
WIDTH approximation and the ABO theory yield
almost identical values.
%
%

\subsection{Atmospheric parameters}

The adopted atmospheric parameters for our programme stars are
given in Table \ref{tabatm} and were taken from
\citet{Pasquini04} and \citet{Pasquini07}.
For the giant star Cl* NGC 6752 YGN 30
we adopted the atmospheric parameters of \citet{Yong}. 
For the two
giants in NGC 6397 we adopted the atmospheric
parameters of \citet{gratton82}. 
For the reader's
convenience the atmospheric parameters of the
giant stars are provided here  on-line
in  Table \ref{gratton}. 
\subsection{Model atmospheres and spectrum synthesis.}
For each star we computed a 1D model atmosphere 
using version 9 of the ATLAS code 
\citep{1993KurCD..13.....K,2005MSAIS...8...14K}
under Linux \citep{2004MSAIS...5...93S,2005MSAIS...8...61S}. 
We used the opacity distribution functions
described by \citet{2003IAUS..210P.A20C} and microturbulent
velocity 1 \kms, the mixing-length parameter, \mlp, was set to 1.25, and the
overshooting switched off.
This model
atmosphere was used as input to the SYNTHE code
\citep{1993KurCD..18.....K,2005MSAIS...8...14K}, 
with different Cu abundances,  
to compute a curve-of-growth for each line. The Cu abundances
were derived by interpolating in these curves of growth.
The corresponding abundances are given in the
second column of Table \ref{abun}, the $\sigma$ is the
variance of the abundances of the two lines. 
The abundances for the individual lines can be found
on-line in  Table \ref{labun}.

The use of three dimensional hydrodynamical
simulations to describe stellar atmospheres 
(hereafter 3D models)
has led to the important notion that 
the outer layers present steeper temperature gradients
than predicted by traditional 1D static model 
atmospheres \citep{A99,A05,jonay} and that
this effect is considerably more pronounced 
for metal-poor stars. In addition to the different
mean temperature profile, the 3D models differ from
traditional 1D models because they account for the
horizontal temperature fluctuations.
Both effects may or may not be important, depending
on the line formation properties of the transition
under consideration.
In order to investigate these effects for the \ion{Cu}{i}
lines we  used  several 3D models computed
with the code \cobold\
\citep{Freytag2002AN....323..213F,Freytag2003CO5BOLD-Manual,Wedemeyer2004A&A...414.1121W}.
The characteristics of the 3D models employed
in this study are given in Table \ref{models}.
The line formation computations for the 3D models
were performed with the \linfor\ 
code\footnote{http://www.aip.de/$\sim$mst/Linfor3D/linfor\_3D\_manual.pdf}.
For each 3D model we used
also  two reference 1D models: the \mD\ and the \xx,
which we define below.

The \mD\ models are computed on-the-fly by
\linfor\ by averaging the 3D model 
over surfaces of equal Rosseland optical depth
and time.
The \mD\ model has, by construction, the mean
temperature structure of the \cobold\ model,
therefore the difference in abundance 
A(3D)-A(\mD), allows us to single out the effects
caused by temperature fluctuations \citep[see][]{zolfito}.

The \xx\ model is  
a  1D, plane parallel, LTE, static, model atmosphere and
employs the same micro-physics and opacity as the
\cobold\ models; it is computed with the LHD code. 
These  models are our models of choice
to define the ``3D correction'' as A(3D) - A(\xx),
where A is the abundance of any given element.
More details on the LHD models may be found in 
\citet{zolfito} and \citet{solphys}. 

In any given \linfor\ run we made computations also
for the \mD\ model and for a \xx\ model, with the same \teff,
\glog , and
metallicity as the 3D model.

\begin{figure}
\resizebox{\hsize}{!}{\includegraphics[clip=true,angle=0]{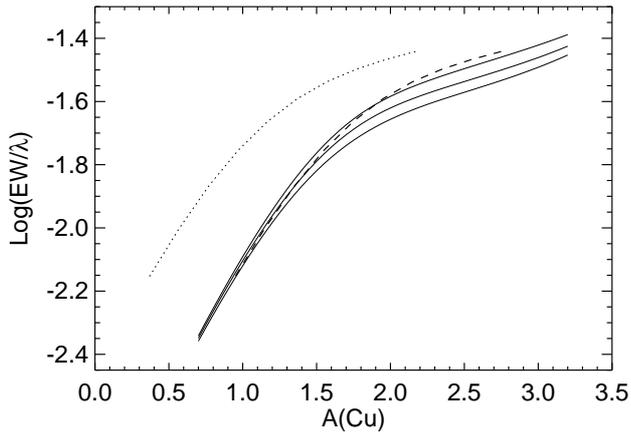}}
\caption{Curves of growth (COG) for the \ion{Cu}{i} 327.3\,nm transition.
The dotted line to the left is the COG for the 3D model
d3t63g40m20n01, the three solid lines to the right
are those for the corresponding \xx\ model for three values
of the microturbulent velocity: 0.5, 1.0 and 1.5 \kms, bottom to top.
The dashed line is the 3D COG shifted arbitrarily by +0.58 dex
along the x-axis. This highlights  that
the shape of the 
3D COG differs from that of the corresponding 1D COGs and therefore
the 3D correction depends on the EW of the transition.
\label{cog}}
\end{figure}

\begin{table}
\caption{Copper abundances for the programme stars.\label{abun}}
\begin{tabular}{lcccc}
\hline\hline\noalign{\smallskip}
Star &     A(Cu)  & $\sigma$ & A(Cu)  & $\sigma$\\
     &  \multicolumn{2}{c}{1D} &\multicolumn{2}{c}{3D}\\
\hline\noalign{\smallskip}
Cl* NGC 6752 GVS  4428  & 3.23 & 0.08 & 2.56 & 0.16\\ 
Cl* NGC 6752 GVS 200613 & 3.01 & 0.05 & 2.23 & 0.07\\
Cl* NGC 6397 ALA 1406   & 1.33 & 0.03 & 0.74 & 0.05\\ 
Cl* NGC 6397 ALA 228    & 1.30 & 0.03 & 0.73 & 0.05\\ 
Cl* NGC 6397 ALA 2111   & 1.19 & 0.02 & 0.60 & 0.02\\ 
HD 218502       & 1.52 & 0.09 & 0.95 & 0.04\\
\hline
\end{tabular}
\end{table}

\begin{table*}
\caption{\cobold\ models employed in the study.}
\label{models}
\begin{center}
\begin{tabular}{ccccccccc}
\noalign{\smallskip}\hline\noalign{\smallskip}
Model & \teff & \glog & [M/H] &\Nt & time & \tchar & Resolution & Box Size  \\
      & K     &       &       &    &  s   &  s     &            & ${\rm Mm}^3$\\
\noalign{\smallskip}\hline\noalign{\smallskip}
d3t50g25mm10n01 & 4990 & 2.5 & $-1.0$ & 20 & 475990 & 1411.9 & $160\times 160\times 200$ & $573.2\times 573.2\times 245.4$\\
d3t50g25mm20n01 & 5020 & 2.5 & $-2.0$ & 20 & 403990 & 1388.3 & $160\times 160\times 200$ & $584.0\times 584.0\times 245.4$ \\
d3t63g40mm10n01 & 6260 & 4.0 & $-1.0$ & 20 & 43800  & 12.2   & $140\times 140\times 150$ & $26.0\times 26.0\times 12.8$\\
d3t63g40mm20n01 & 6280 & 4.0 & $-2.0$ & 16 & 27600  & 49.0   & $140\times 140\times 150$ & $26.1\times 26.1\times 12.8$\\
d3t63g45mm10n01 & 6240 & 4.5 & $-1.0$ & 20 & 24960  & 16.0   & $140\times 140\times 150$ & $7.0\times 7.0\times 4.0$\\
d3t63g45mm20n01 & 6320 & 4.5 & $-2.0$ & 19 & 9120   & 15.9   & $140\times 140\times 150$ & $7.0\times 7.0\times 4.0$\\
\noalign{\smallskip}\hline\noalign{\smallskip}
\end{tabular}
\end{center}
\end{table*}

\begin{table}
\caption{Mean copper abundances for the two clusters.\label{meanabu}}
\begin{tabular}{lcccc}
\hline\hline\noalign{\smallskip}
Star &     A(Cu)  & $\sigma$ & A(Cu)  & $\sigma$\\
     &  \multicolumn{2}{c}{1D} &\multicolumn{2}{c}{3D}\\
\hline\noalign{\smallskip}
NGC 6752 dwarfs & 3.04 & 0.07 & 2.28 & 0.12 \\  
NGC 6752 giants & 2.03 & 0.05 & 1.98 & 0.05 \\
NGC 6397 dwarfs & 1.25 & 0.05 & 0.63 & 0.04 \\
NGC 6397 giants & 1.40 & 0.17 & 1.30 & 0.17 \\ 
\hline
\end{tabular}
\end{table}

The computation of a 3D model is still very time
consuming, even on modern computers (several months), 
it would be impractical to compute a specific 3D model
for any set of our atmospheric parameters.
Our strategy is therefore the following: we perform
an abundance analysis with ATLAS model atmospheres 
computed for the desired set of atmospheric parameters;
we use a grid of 3D models with atmospheric parameters
that bracket the desired ones and compute the
relevant 3D corrections by linear or bi-linear
interpolation in the grid, as appropriate; the
3D abundance is obtained by applying the interpolated
3D correction to the 1D abundance.

The line formation computations were performed
using SYNTHE for the ATLAS models and \linfor\
for all  other models. As a consistency check we
used \linfor\ with an ATLAS model as input and verified
that the line profiles and EWs are consistent with
the results derived from SYNTHE+ATLAS.
The difference between the two line formation codes
amounts to a few hundredths of dex in terms of abundance, 
a quantity that is
irrelevant with respect to the size of the 3D corrections
under consideration.

For three out of the six TO stars under study,
the \ion{Cu}{i} lines are strong (EW$ > 4.0$\,pm),
therefore they are surely  in the saturation
regime. Their 3D correction depends
on the adopted microturbulence in the adopted reference
1D atmosphere. 
To take this into account, different \xx\ curves
of growth were computed with microturbulent
velocities of 0.5, 1.0 and 1.5 \kms, to allow us
interpolation to any desired value of $\xi$.
For weaker lines the microturbulent velocity does not
play a fundamental rule, so that the 3D correction is mostly
insensitive on the choice of this parameter. 
We find that the 3D curve of growth is not
a simple translation of a 1D curve of growth, 
but  has a distinct shape. An example 
to illustrate the effect 
is shown in  Fig.\,\ref{cog}.
An immediate consequence is that for all lines we considered,
the 3D correction depends on the EW, even for the weaker lines. 
 We note that  the 
dependence of the
3D correction on the microturbulence
implies that the  abundance
obtained by applying the correction to an abundance
derived from a 1D model
depends on the adopted microturbulence.
One of the reasons to prefer the use of 3D models
is to avoid the use of this parameter.
The correct way to treat this is, not to 
use the 3D correction, but to derive the abundance
by an interpolation in a set of 3D curves of
growth, or to use suitable fitting functions, 
as done for lithium by \citet{sbordone_li}.
But this requires the use of a larger set of
3D models, which bracket the effective temperatures
of the studied stars. For the purpose of the present
exploratory investigation we believe the approach
of using 3D corrections is adequate, especially because 
the conclusion of the study is that 3D-LTE abundances
are unreliable. Finally we  point  
out that it may be that current 3D models
do not correctly capture turbulence at small
scales \citep{Steffen}.
If this were indeed the case all  abundances
derived from saturated lines are doubtful,
whether they are derived applying a 3D correction
or directly derived from the 3D curves of growth.

We therefore computed the 3D correction 
for each of the measured EWs for each of
the relevant 3D models.
In general the 3D correction
will also depend on the treatment
of convection in the 1D reference model, 
hence on the adopted \mlp.
The lines under consideration do not form
in the deepest layers, which are the most
affected by the choice of \mlp, thus
they are insensitive to it.
All  \xx\ models employed have \mlp = 1.0.
The computed A(3D) -- A(\xx) corrections,
as well as the A(3D) -- A(\mD) corrections
are given on-line in Table \ref{corr3D}.
The 3D abundances provided in Table \ref{abun}
are obtained by applying to the 1D abundances the
3D corrections in Table \ref{labun},
which were obtained by interpolating the corrections
in Table \ref{corr3D}.

The six stars under study have very similar
effective temperatures. All are within
roughly 100 K of the effective temperatures of
the 3D models listed in 
Table \ref{models}
(\teff $\sim 6300$ K). Therefore
it is not necessary to include more 
3D models and perform an interpolation 
in \teff.
The metallicities and gravities of the models
in Table \ref{models} bracket the 
metallicities and gravities in Table \ref{tabatm}.
We used a bi-linear interpolation in metallicity
and gravity for the two stars of NGC 6752 and for
HD 218502. The three stars of NGC 6397, though 
have a metallicity of almost --2.0, therefore
a linear interpolation in surface gravity
was sufficient.

The computation of 3D models for typical giant stars is much more time
consuming than for F-type and cooler dwarfs. The radiative relaxation time in the surface
layers of warm giants becomes significantly shorter than the dynamical time
scale, which makes 
it computationally expensive to properly capture the time
evolution of the system.
At present we  have only two fully relaxed 
models of  metal-poor giants.  The parameters
are given in Table \ref{models} and the metallicities
are --1.0 and --2.0.
The surface gravity is larger than that of the 
majority of the giant
stars analysed in either cluster. Only the faintest
stars analysed by \citet{Yong} have parameters
\teff\ and \glog\ close to  those of our giant
models. 
In spite of this we believe that we can use the   3D corrections
derived from this model, as a first
order approximation as representative of 
the corrections in giant stars of both clusters.
This is possible because the 3D corrections are rather small, 
especially when compared to those of dwarf stars.
This is partly because giant 3D models
do not show  a pronounced over-cooling, compared 
to 1D models, as dwarfs display.
This conclusion is also based on the examination
of several snapshots of not fully relaxed giants
of different atmospheric parameters, which we are in the
process of computing.
The 3D correction for the giant 
models is --0.1 dex for both the examined \ion{Cu}{i}
lines of Mult.\,2 for the model of metallicity --2.0 
and 0.0 dex (actually +0.01) for the model
of metallicity --1.0. 
We apply a correction of --0.1 to the abundances
of  giant stars  in NGC 6397 and 
--0.05 to those of NGC 6752.

\begin{figure}
\resizebox{\hsize}{!}{\includegraphics[clip=true,angle=0]{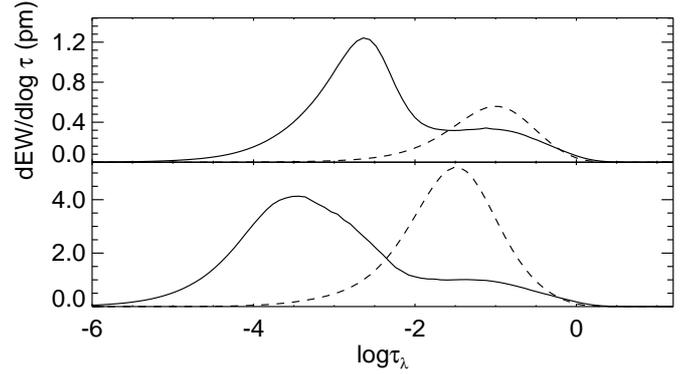}}
\caption{Contribution functions  of the EW at disc-centre,
defined in a way that their integral over $\log\tau_\lambda$
gives the EW \citep{magain86}, for the \ion{Cu}{i}
324.7\,nm line and the model d3t63g40mm20n01 for two different
values of Cu abundance. In the upper panel A(Cu)=0.2, in the
lower panel A(Cu)=1.7. The solid lines refer to the 3D model,
the dashed lines to the corresponding \xx\ model. \label{contf}}
\end{figure}

\begin{figure}
\resizebox{\hsize}{!}{\includegraphics[clip=true,angle=0]{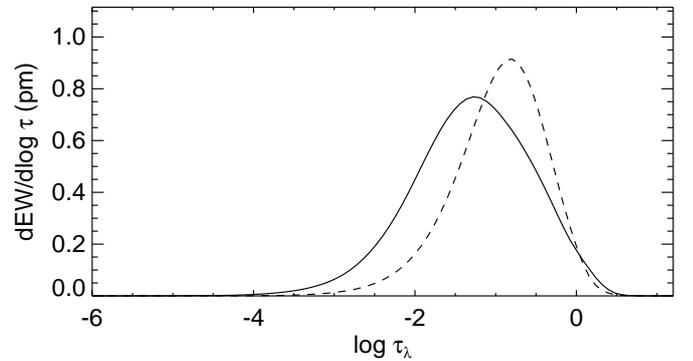}}
\caption{Contribution functions for the \ion{Cu}{i}
510.5\,nm line and the model d3t50g25mm20n01. 
The solid line refers to the 3D model,
the dashed line to the corresponding \xx\ model. \label{cf5105}}
\end{figure}

\section{Results}

\subsection{HD 218502}
The analysis of the reference star HD 218502
shows that the abundances derived from the two
\ion{Cu}{i} resonance lines (Table \ref{labun}) 
agree well,
both with 1D and 3D models. This gives us 
confidence in the reliability of the atomic
data used. It also 
suggests that with good quality
data, the EWs of both lines can be satisfactorily
measured in spite of the complexity of the
spectral region, especially for the 324.7\,nm line.
Our 1D abundance (A(Cu)$=1.52\pm 0.09$) agrees,
within errors with what reported 
by \citet[][A(Cu)$ = 1.70 \pm 0.17$]{Bihain}.
We note a small difference in the effective temperature
adopted in the two analyses (about 100 K) and a difference
in the data: our UVES spectra are of considerably higher
quality than the CASPEC spectra used by \citet{Bihain},
who measured only the 327.3\,nm line.  
This agreement is expected because the atomic data
are the same in the two analyses and the 1D model atmospheres
used are similar, the main difference being the overshooting.
This  check suggests that we may reasonably 
assume that our Cu abundances should be consistent with
those of  \citet{Bihain}, which are based on the UV resonance 
lines.
An inspection of Fig. 6 of \citet{Bihain}
suggests that these measurements  substantially agree
with those of \citet{Mishenina}, which are essentially based
on the measurements of the lines of Mult.\,2. Yet 
one should take into account the large error bars that derive
from the relatively poor S/N ratio that is achievable in the UV range.

The 3D correction for the lines of Mult.1 is  large
and the 3D abundance in this star is well below
all the measurements in giant stars of similar
metallicity.
An application of 3D corrections to all  measurements
of \citet{Bihain} would probably break the agreement
with the measurements of \citet{Mishenina}.
We have inspected the available red spectra
of HD 218502, to see if any of the lines
of  Mult.\,2 could be detected, but this was
not the case. 

\subsection{TO stars in NGC 6752 and NGC 6397}

For the cluster stars 
there is also a good consistency between the abundances
derived from the two resonance lines for
any given line, in spite of the
much lower S/N ratios in the cluster stars.
This suggests that there is no major
inconsistency in the EW measurements.

The weighted mean\footnote{
The error on the weighted mean has been taken to be
the largest between $\sqrt{\left(\sum{1\over\sigma_i^2}\right)^{-1}}$
and $\sqrt{\left(\sum{1\over\sigma_i^2}\right)^{-1}\times{1\over (n-1)}
\sum{\left(x_i-<x>\right)^2\over\sigma_i^2}}$, where $x_i$ are the data
points $\sigma_i$ are the associated errors,  $\left\langle x\right\rangle$ is
the mean value and $n$ is the number of data point
\citep[see e.g.,][pages 144--150]{agekan}.}
of the abundances
of the clusters, reported in Table \ref{meanabu},
displays an error in the mean that is reasonably 
small. We believe  that the mean abundances
for the two clusters obtained from the dwarf
stars are indeed representative of the 
\ion{Cu}{i} abundance derived from the 
lines of Mult.\,1.
Considering that each spectrum of a
TO star in these clusters amounts to  about 10 hours
of integration with UVES, it is unlikely
that in the near future better data
or data for a larger number of stars
will be available, although it would
clearly be desirable.

\subsection{Star  Cl* NGC 6752 YGN 30}

In this star, like in the other giant stars
observed by \citet{Yong}, both the lines
of Mult.\,1 and of Mult.\,2 are measurable,
which provides for a consistency check.
We did not measure the line at 324.7\,nm 
since the region is extremely crowded in these
cool giants, but the line at 327.3\,nm 
is clean and unblended. For Mult.\,2
we only measured the 510.5\,nm line, 
since the other line is not present in the spectrum,
because it falls in the gap between the two CCDs.
The two lines (324.7\,nm and 510.5\,nm )
provide inconsistent results, the line of Mult.\,1
provides an abundance that is 0.54 dex higher
than that of the line of Mult.\,2 (see on-line Table \ref{labun}).
The abundance we derive from the line of Mult.\,2
substantiallly agrees with the measure
of \citet{Yong}, our abundance is 0.13 dex smaller.
Of this difference 0.05\,dex are because 
we use ATLAS models without overshooting,
while  \citet{Yong} use models
with overshooting, the remaining difference should 
be attributed to a difference in the measured EW and
possibly to the different spectrum synthesis
codes used (MOOG by \citealt{Yong} and SYNTHE by us).
The 3D corrections for both  lines of Mult.\,1 
and Mult.\,2 are small and agree to within a few hundredths
of dex. The abundance
from the two lines cannot be brought into agreement
by using 3D models.

\subsection{Giant stars in NGC 6397}
The error in the mean of the two giants is
0.18 dex, which is essentially identical
to the estimate of \citet{gratton82} of 0.2 dex
on the copper abundance.
Even making use of 1D models the abundance of
the giant stars is considerably higher
than in the dwarf stars.

\section{Effects of atmospheric parameters\label{atmpar}}

We intend to quantify the effect of changing atmospheric
parameters on the derived abundances.
The abundances derived from the \ion{Cu}{i} resonance lines
are fairly sensitive, for a neutral species, to
the adopted surface gravity.
For dwarf stars a change of $\pm 0.25$\, dex in \glog\
induces a change of $\mp 0.1$\,dex in abundance.
For the giant stars they are only slightly less sensitive,
$\mp 0.06$\,dex. On the other hand for giant stars the dependence
on gravity of the abundances derived from the lines
of Mult.\,2 is very weak $\mp 0.01$\,dex.
An inspection of on-line  Table \ref{corr3D} allows 
us to estimate
the effect of surface gravity on the 3D corrections
for the lines of Mult.\,1.
By increasing the gravity by 0.5\,dex, the 
3D correction increases by 0.1 to 0.2 dex, depending
on how saturated the line is. Since the 3D correction
is negative, this means that it decreases in absolute value.
The opposite trends with surface gravity on the 1D abundance
and 3D correction imply that the two effects tend to cancel
and the overall sensitivity of
the 3D abundance on surface gravity is small.

The dependence of abundances on effective temperatures 
for the lines of Mult.\, 1 is similar for dwarfs and
giants, and is about $\pm 0.2$\,dex for  
a change of $\pm 100$ K in effective temperature.
To evaluate the dependence of the 3D corrections
on the effective temperature we used four models,
extracted from the CIFIST grid \citep{ludwigJD10}.
All four models have
effective temperature around 5900\,K, 
their metallicities are
--1.0 and --2.0 and their \glog\ 4.0 and 4.5.
The result is that for a decrease of 300\, K 
the 3D correction increases by 0.2\,dex.
Again the variation of the 3D correction goes in the 
opposite direction with respect to the variation 
in the 1D abundance. Combining the results we
conclude that a decrease of 300\,K in effective
temperature results in a decrease by 0.4\,dex
in copper abundance (3D). 

For the giants stars we also estimated the variation 
of the abundances derived from the lines of
Mult.\, 2, which amounts to about $\pm 0.1$\,dex
for a variation of $\pm 100$\,K in effective temperature.

\section{Hydrodynamical models and 
spectrum synthesis.}

Clearly the large 3D corrections derived for the 
\ion{Cu}{i} resonance lines
are driven by the fact that in the outer layers the 
hydrodynamical models are considerably cooler than the
corresponding 1D models, the so-called ``overcooling''.
This shifts the ionisation equilibrium towards neutral 
copper, and as a consequence the line contribution
function shows  a strong peak in these layers,
contrary to the 1D model (se Fig.\,\ref{contf}).
From a physical point of view this is expected,
simply because the hydrodynamical
model is observed to  transport flux through
convection even in layers where the corresponding
1D model is formally stable against convection (overshooting). 
One should however ask to which extent  the computed overcooling
depends on the assumptions made.
\citet{BonifacioRio} pointed out the difference in the
overcooling for metal-poor giants computed with \cobold\ and
that computed by \citet{collet}.
\cite{BeharaRio} pointed out that for extremely metal-poor dwarfs
the overcooling is considerably less in \cobold\ models computed
with 12 opacity bins than in models computed with
6 opacity bins, like the ones used here.

\begin{figure}
\resizebox{\hsize}{!}{\includegraphics[clip=true,angle=0]{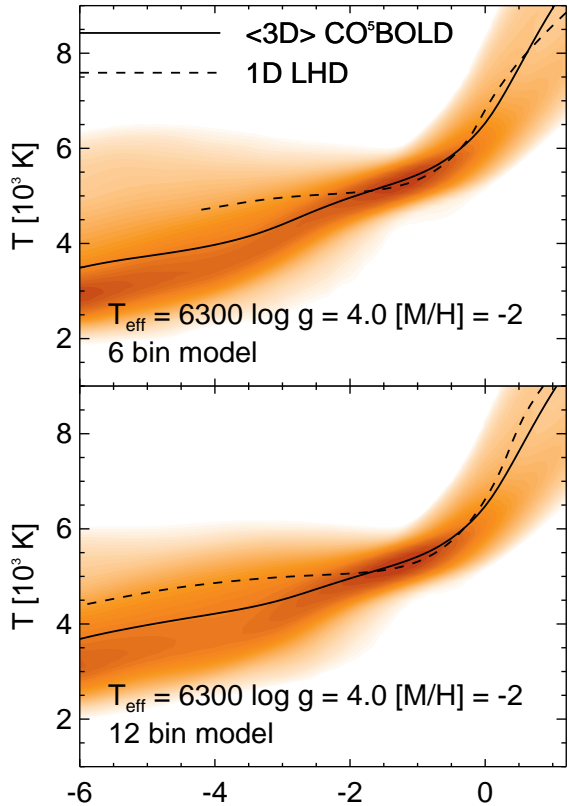}}
\caption{Comparison between the
temperature structures of two models with \teff = 6300, \glog = 4.0
and metallicity --2.0, computed with 6 and 12 opacity difference.
The difference is not as large as it is at lower metallicities
(see Fig.\,1 of \citealt{BeharaRio}). 
\label{T_tau}}
\end{figure}

\begin{figure}
\resizebox{\hsize}{!}{\includegraphics[clip=true,angle=0]{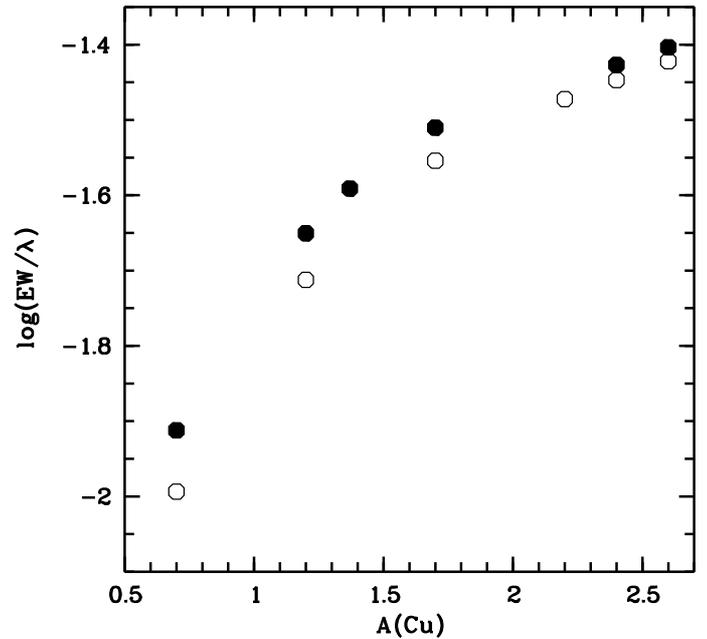}}
\caption{Comparison between the
curves of growth for the 327.3\,nm line computed for
the two models shown in Fig.\,\ref{T_tau}. 
Solid symbols correspond to the 6 bin model and open symbols
to the 12 bin model.\label{cog6_12}}
\end{figure}

For the time being we have relatively few \cobold\ models computed
with 12 opacity bins, among which we have one with parameters
identical to one used in the present investigation: \teff = 6300,
\glog = 4.0 and metallicity = --2.0.
In  Fig.\,\ref{T_tau} we show the mean temperature structures
of the two models. Obviously the difference
is smaller than what is displayed by the models 1\,dex more
metal-poor shown by \citet{BeharaRio}.
The qualitative conclusion is confirmed by comparison of the
curves of growth. In Fig.\,\ref{cog6_12} the curve of growth
for the 327.3\,nm line used in the present investigation 
is compared to the one computed from the corresponding
12 bin model. As can be appreciated from the plot, 
for a given equivalent width the 12 bin model will yield
a Cu abundance that is higher by approximately 0.1\,dex, thus
correspondingly decreasing the 3D correction.

Another matter of concern is that the current
version of \linfor\ treats scattering as true absorption.
That is to say,  although scattering processes such as
Rayleigh scattering off hydrogen atoms are taken into 
account as opacity sources, in the solution of the transfer
equation the source function is set equal to the local
Planck function, without a term depending on the mean
radiation intensity ($S_\nu = B_\nu$).
To which extent can this approximation affect our computations,
especially in the near ultraviolet, where scattering 
processes are a non-negligible source of opacity\,?
We  assessed this by using 1D models and 1D spectrum
synthesis. We used a slightly
modified version of the SPECTRV code in the SYNTHE suite 
so that  scattering is treated 
as true absorption when the card SCATTERING OFF is set in
the input model (see e.g. \citealt{Castelli88}).  
We computed
line profiles both with SCATTERING ON and SCATTERING OFF
for the model at \teff = 6296 K \glog = 4.0 and metallicity --2.0
(rlevant to HD 218502) and for the
model \teff = 4943 \glog = 2.42 and metallicity --1.5
(relevant to Cl* NGC 6752 YGN 30).
It turns out that in both cases the difference is irrelevant
to our analysis
(0.6\% in the continuum and 0.03\% in the residual intensity
for the dwarf model and 14\% in the continuum and 2\% in the
residual intensity).
The effect of treating scattering as true absorption
is very similar on the continuum and in the lines, thus
implying a small effect on the residual intensity and equivalent width.

Recently \citet{hayek} have introduced a proper treatment
of scattering in their magneto-hydrodynamical simulation code,
BIFROST.
For the Sun and solar-type stars they do not find a 
significant impact of continuum scattering  on the temperature structure. 
We thus believe that although the impact of a proper
treatment of scattering needs to be investigated in metal-poor
dwarfs, it seems unlikely that the results presented here 
will be seriously challenged by this.

\section{Discussion}

If we consider the abundances  
in Table \ref{meanabu} at face value we are 
led to the inescapable conclusion that 
the Cu abundances in dwarfs and giants
do not agree. 
Even though the abundances
are almost compatible, within errors, at least in the 1D case,
this systematic difference should not
be overlooked.
The behaviour is different
between the two clusters: in NGC 6752 the dwarfs provide
a {\em higher} abundance than the giants, while in the
case of NGC 6397, they provide a {\em lower} value than the
giants; this both using 1D and 3D models.
For NGC 6752 the difference in 1D abundances is 
of one order of magnitude (1\,dex), but this
is reduced to only  0.3\,dex if we look at the
3D abundances, which are almost compatible with the errors.
The situation is reversed in NGC 6397, in 1D the 
abundances of dwarfs is only 0.15\, dex 
lower than that in giants, while
in 3D, the diferece is 0.9\, dex.
This behaviour may be understood in 
terms of the different line formation properties
in dwarfs and giants and how they change with different
Cu abundance.
In Fig.\,\ref{contf} we show, as an example,
the contribution functions of the EW at disc-centre for one of our models
for a dwarf star, for two different Cu abundances.
In the top panel A(Cu)=0.2 and the line is on the linear
part of the curve of growth, in the bottom panel,
A(Cu)=1.7 and the line is saturated.
In both cases the 3D contribution function is 
very different from the 1D one and is peaked in the outer
layers of the atmosphere.
The formation of the lines of Mult.\,2 \relax in the atmospheres
of giants is instead much less affected by 3D effects,
as depicted in Fig.\,\ref{cf5105}.
The contribution functions of the lines of Mult.\,1 \relax in
the giant model are morphologically similar to
what is shown in Fig.\,\ref{cf5105}, confirming 
the weak overcooling present in this model.
While the above arguments explain the behaviour of the 
3D corrections, they have no bearing in the abundance
difference that we find between dwarfs and giants.

In these situations one has always to consider two
possible alternatives: i) the difference is true and has
an astrophysical origin; ii) the difference arises from
shortcomings in the analysis.
In our opinion hypothesis i) must be discarded.
It seems extremely far fetched 
to devise a physical mechanism by which Cu should
be overabundant in dwarf stars, like in NGC 6752,
while the reverse is true in NGC 6397. 
Where giants
have a higher Cu abundance, one could imagine
to explain such a scenario either by invoking
diffusion in TO stars, or Cu production 
in giant stars, or perhaps even a combination
of both. However, an examination of Table
\ref{meanabu} shows that the differences
that need explanation are far too large
to be created by diffusion, and the Cu production
would also have to be highly efficient. 
In addition note that the Cu abundances
in the large sample of \citet{Yong} 
are extremely uniform, which 
speaks against 
Cu production.

Thus we are left with the conclusion that 
the abundance determinations in either dwarfs
or giants, or both, are wrong.
Let us start by examining the 3D abundances in 
dwarf stars. Contribution functions like those
shown in Fig.\,\ref{contf} must give rise to concerns
about the LTE approximation used in our computations.
The outer and less dense layers of the atmosphere, 
which contribute mostly to the line EW, are those
in which the photon mean free path is longest
and deviations from LTE may be expected. 
The situation is even more extreme in a 3D atmosphere
where photons from a hot up-draft may
transfer horizontally and overionise a neighbouring
cool down-draft.
A morphologically similar situation is indeed observed
for the \ion{Li}{i}  doublet in metal-poor stars \citep{A03,cayrel}.
In LTE the contribution function displays a double peak,
with a substantial contribution from outer cool layers, in NLTE
this peak is entirely suppressed by overionisation. 
The nearly exact cancellation between 3D and NLTE correction
that takes place for the  \ion{Li}{i}  doublet, and results
in 3D-NLTE abundances in very close agreement with
1D-LTE abundances, should not be taken as a general rule.
Nevertheless we believe that the \ion{Cu}{i} lines of Mult.\,1
cannot be described by 3D-LTE computations, but NLTE effects
should be properly accounted for.

This leads to  the question of how reliable the LTE
approximation is for the 1D computations.
That even in 1D the abundances in 
dwarfs and giants differ
by a factor of the order of 2, surely prompts to see if
for NLTE computations the two sets of abundances
may be brought into agreement.
\citet{Bihain} tried to fit the \ion{Cu}{i} lines of Mult.\,1
in the solar spectrum, but were unable to reproduce the core of
the lines. This was attributed mainly to the presence
of a chromosphere that influences the cores of strong lines.
Deviations from LTE, however, could be another, possibly concomitant,
cause for the failure to reproduce the line core in LTE.
For Pop II dwarf stars  
chromospheric effects should not
be strong, in view of their old age,
even if chromospheres were present.

Let us finally consider the Cu abundances in giant stars.
Our limited computations suggest that they should not suffer
large 3D effects. 
The test we conducted for the giant star in NGC 6752
suggests that the LTE synthesis does not allow us
to reproduce satisfactorily the lines of Mult.\,1
and of Mult.\,2 with the same Cu abundance.
Indeed  the discrepancy is quite large (0.5 dex);
significant deviations from LTE for either or both sets
of lines could be responsible for this.
Which should be further investigated in order to produce
reliable Cu abundances. The evolution of copper with metallicity,
essentially based on the measurements of Mult.\,2 \relax in giant stars, 
shows a rather sharp drop in [Cu/Fe] around [Fe/H]=--1.5.
This means that it takes place around A(Cu)=2.0.
In the curve of growth for the 510.5\,nm line of Mult.\,2 for our giant
model, this is roughly the abundance for which the line begins
to enter in the saturation regime. 
If the \ion{Cu}{i} lines of Mult.\,2 suffer deviations
from LTE it is likely that these depend on the line strength
and may show a rather sharp change just when the line
enters a saturation regime. 
This behaviour is observed, for instance,  for the sodium D lines 
\citep[see Fig. 6 of][]{andrievsky}.
These considerations render NLTE computations
for Cu very desirable.

Unfortunately, to our knowledge, up to now no such computations
have been published, nor does a Cu model atom exist. 

Among the possible causes for the discrepancy in abundances between
giants and dwarfs one may  also consider 
errors in the atmospheric parameters.
One may conclude  that this cannot be the case
by noticing the discrepancy between the Cu abundance
derived from Mult.\,1 and Mult.\,2 \relax in 
star Cl* NGC 6752 YGN 30, for which lines of both
multiplets are measured. Given that the response
of both multiplets to a change in effective temperature 
is similar, the conflicting results cannot
be resolved by changing the effective temperature.
Of course the discrepancy between the two
multiplets can be resolved in 1D by invoking a higher 
microtrubulence, but it would be necessary
to raise it by 1\,\kms . This increase would
then cause a strong trend between iron abundances
and equivalent widths. Furthermore this would not
allow us to solve the discrepancy in the 3D analysis.
Indeed while the 3D correction for Mult.\,2 would not change
significantly with this increase in microturbulence,
those of Mult.\,1 would increase by about 0.4\,dex, thus breaking
the agreement between the two multiplets forced in 1D.
Formally one can certainly find a value of the 
microturbulence
that forces the  1D abundance plus
3D correction   of the two mutliplets to be equal, 
while leaving a discrepancy in the 1D abundances.
This however would again cause an abundance spread
among lines of other elements (e.g. iron) and 
can hardly be invoked as a solution of the problem.
Although at the moment we do not have enough 3D models
for giant stars to perform a full 3D analysis,
as done by \citet{sbordone_li} for lithium, we believe 
that our results indicate that this analysis will
provide discrepant abundances from the two multiplets.

Let us further consider if changes in the atmospheric
parameters of giant or dwarf stars in either cluster
may allow us to reconcile their copper abundances.
As discussed in Sect.\,\ref{atmpar} a decrease in effective 
temperature of 300\,K in dwarf stars implies a decrease 
by about 0.4\,dex. Therefore for the dwarf stars in NGC 6752
a decrease in effective temperature by about 225\,K
while keeping the temperature of giants constant, 
would reconcile the copper abundances of the to sets
of stars. While not implausible, this change would certainly
cause a mismatch in the abundances of other elements
between giants and dwarfs, most notably iron, which would become
less abundant in dwarfs than in giants.
While one could argue that atmospheric phenomena such as diffusion
may alter abundances of dwarf stars, it seems then contrived
to invoke a different behaviour between copper and iron.
But let us now turn to the other cluster, NGC 6397. 
Here the situation is reversed, the dwarfs display a lower
abundance than giants. However here one would need
to invoke an increase in effective temperatures of the dwarf
stars by over 500 K, placing them at \teff\ around 6700\,K
and the cluster turn-off at about 6900\,K.
While these exceedingly high temperatures may be appealing
(one would immediately solve the cosmological lithium problem!)
they appear impossible to reconcile with the colours of the 
cluster and theoretical isochrones. 

Although the precise value of the
atmospheric parameters assigned to the stars
certainly plays a role in the difference in copper
abundances between dwarfs and giants in the two clusters,
we may dismiss the hypothesis that it may be cancelled by a 
suitable choice of parameters.

\section{Conclusions}

Our study of the \ion{Cu}{i} lines of Mult.\,1 
in the TO stars of the globular clusters NGC 6397
and NGC 6752 allow us to draw several conclusions:

\begin{enumerate}
\item
the Cu abundance derived from the  \ion{Cu}{i} lines
of Mult.\,1 \relax in dwarf stars differs
from that derived in giants from the \ion{Cu}{i}
lines of Mult.\,2 \relax in the same clusters;
\item 
the  \ion{Cu}{i} lines
of Mult.\,1 \relax in dwarf stars
show large 3D corrections when computed in LTE;
\item
the  \ion{Cu}{i} lines
of Mult.\,2 \relax in giant stars
show small 3D corrections when computed in LTE;
\item
the contribution functions
of  the  \ion{Cu}{i} lines
of Mult.\,1 \relax in dwarf stars
suggest that these cannot be reliably
computed under the LTE approximation;
\item
for the only star for which we have measured
both the lines of Mult.\,1 and of Mult.\,2
we find that the derived abundances disagree 
by 0.5 dex, both using 1D and 3D models.
\end{enumerate}

From the above we conclude that 
the  \ion{Cu}{i}
lines of Mult.\,1 \relax 
are not reliable indicators
of copper abundances. A full 3D-NLTE treatment
of these lines should be used.
When treated in 1D these lines yield abundances
which, in the studied cases, differ from the
abundances in giants by 0.2 to 1\,dex, 
but the difference can be in either direction
(lower or higher abundance in dwarfs).
That the dwarf/giant discrepancy is in
opposite directions for two clusters that differ 
by only 0.5\,dex in metallicities also suggests that
NLTE alone might not be able to reconcile
the abundances in the two groups of stars.
Whether the combination of 3D and NLTE effects
may achieve this is an open issue, although
it seems to be doubtful. Clearly a new investigation
of Cu abundances in giants of NGC 6397 would be
highly welcome, to place this discussion on firmer grounds. 

Investigation of departures from LTE both
in dwarfs and giants is needed to place the
copper abundances on a firm footing.
Without knowledge of these departures we recommend
that the abundances measured in giants from 
lines of Mult.\,2 should be preferred 
for the studies of chemical evolution. 
The main motivation for this recommendation is that the 3D effects
for the \ion{Cu}{i} lines
used in giants are rather small.

\begin{acknowledgements}
  We are grateful to L. Pasquini for many useful
  comments on an early version of this paper.
  The authors P.B., E.C., H.-G.L. acknowledge financial
  support from EU contract MEXT-CT-2004-014265 (CIFIST).
  We acknowledge use of the supercomputing centre CINECA,
  which has granted us time to compute part of the hydrodynamical
  models used in this investigation, through the INAF-CINECA
  agreement 2006, 2007.
\end{acknowledgements}

\bibliographystyle{aa}

\Online

\begin{table}
\caption{Atomic data for the \ion{Cu}{i} lines used.\label{tabgf}}
\begin{tabular}{lccc}
\hline\hline\noalign{\smallskip} 
Isotope & $\lambda$ &\gflog & $\chi$\\
        &  nm       &       &  eV   \\
\hline\noalign{\smallskip}
$^{65}$Cu  & 324.7510&$-0.868$ & 0.000\\
$^{63}$Cu  & 324.7512&$-0.868$ & 0.000\\
$^{65}$Cu  & 324.7512&$-0.868$ & 0.000\\
$^{63}$Cu  & 324.7513&$-0.868$ & 0.000\\
$^{65}$Cu  & 324.7513&$-1.266$ & 0.000\\
$^{63}$Cu  & 324.7514&$-1.266$ & 0.000\\
$^{63}$Cu  & 324.7551&$-0.421$ & 0.000\\
$^{65}$Cu  & 324.7552&$-0.421$ & 0.000\\
$^{63}$Cu  & 324.7553&$-0.868$ & 0.000\\
$^{63}$Cu  & 324.7554&$-1.567$ & 0.000\\
$^{65}$Cu  & 324.7554&$-0.868$ & 0.000\\
$^{65}$Cu  & 324.7555&$-1.567$ & 0.000\\
\\
$^{63}$Cu  & 327.3927&$-0.864$ & 0.000\\
$^{63}$Cu  & 327.3930&$-1.563$ & 0.000\\
$^{63}$Cu  & 327.3969&$-0.864$ & 0.000\\
$^{63}$Cu  & 327.3972&$-0.864$ & 0.000\\
$^{65}$Cu  & 327.3925&$-0.864$ & 0.000\\
$^{65}$Cu  & 327.3929&$-1.563$ & 0.000\\
$^{65}$Cu  & 327.3970&$-0.864$ & 0.000\\
$^{65}$Cu  & 327.3973&$-0.864$ & 0.000\\
\\
$^{65}$Cu  & 510.5503&$-3.720$ & 1.389\\
$^{63}$Cu  & 510.5505&$-3.720$ & 1.389\\
$^{65}$Cu  & 510.5506&$-2.766$ & 1.389\\
$^{63}$Cu  & 510.5509&$-2.766$ & 1.389\\
$^{65}$Cu  & 510.5509&$-2.720$ & 1.389\\
$^{65}$Cu  & 510.5510&$-3.896$ & 1.389\\
$^{63}$Cu  & 510.5511&$-2.720$ & 1.389\\
$^{63}$Cu  & 510.5512&$-3.896$ & 1.389\\
$^{65}$Cu  & 510.5515&$-2.653$ & 1.389\\
$^{63}$Cu  & 510.5517&$-2.653$ & 1.389\\
$^{65}$Cu  & 510.5519&$-2.398$ & 1.389\\
$^{63}$Cu  & 510.5520&$-2.398$ & 1.389\\
$^{65}$Cu  & 510.5530&$-2.750$ & 1.389\\
$^{63}$Cu  & 510.5531&$-2.750$ & 1.389\\
$^{63}$Cu  & 510.5536&$-2.148$ & 1.389\\
$^{65}$Cu  & 510.5536&$-2.148$ & 1.389\\
$^{63}$Cu  & 510.5558&$-1.942$ & 1.389\\
$^{65}$Cu  & 510.5560&$-1.942$ & 1.389\\
\\
$^{65}$Cu  &  578.2053&$-2.924$&1.642 \\
$^{65}$Cu  &  578.2062&$-3.225$&1.642 \\
$^{63}$Cu  &  578.2066&$-3.225$&1.642 \\
$^{65}$Cu  &  578.2074&$-2.526$&1.642 \\
$^{63}$Cu  &  578.2078&$-2.526$&1.642 \\
$^{65}$Cu  &  578.2105&$-2.526$&1.642 \\
$^{63}$Cu  &  578.2106&$-2.526$&1.642 \\
$^{63}$Cu  &  578.2117&$-2.526$&1.642 \\
$^{65}$Cu  &  578.2117&$-2.526$&1.642 \\
$^{65}$Cu  &  578.2173&$-2.079$&1.642 \\
\hline
\end{tabular}
\end{table}

\begin{table*}
\caption{Copper abundances from the individual lines.\label{labun}
}

\begin{tabular}{lccccl}
\hline \hline\noalign{\smallskip}
 Star & Type & Line & EW & A(Cu)$_{1D}$ & 3D-\xx \\
      &  G/D & nm   & pm & dex   &  dex    \\
\hline\noalign{\smallskip}
Cl* NGC 6752 GVS  4428   & D & 324.7 & 12.66 & 3.29 &$ -0.62$ \\ 
Cl* NGC 6752 GVS  4428   &   & 327.3 & 11.31 & 3.17 &$ -0.73$ \\
\\                          
Cl* NGC 6752 GVS 200613  & D & 324.7 & 11.33 & 3.04 &$ -0.76$ \\ 
Cl* NGC 6752 GVS 200613  &   & 327.3 & 10.44 & 2.97 &$ -0.79$ \\
\\                          
Cl* NGC 6752 YGN 30      & G & 327.3 &16.81 & 2.43 &$ -0.05$ \\
Cl* NGC 6752 YGN 30      &   & 510.5 & 2.32 & 1.89 &$ -0.05$ \\
\\
Cl* NGC 6397 ALA 1406    & D & 324.7 & 5.35 & 1.36 &$ -0.58$ \\
Cl* NGC 6397 ALA 1406    &   & 327.3 & 3.44 & 1.31 &$ -0.59$ \\
\\                          
Cl* NGC 6397 ALA 228     & D & 324.7 & 5.34 & 1.28 &$ -0.58$ \\  
Cl* NGC 6397 ALA 228     &   & 327.3 & 3.94 & 1.32 &$ -0.57$ \\  
\\                          
Cl* NGC 6397 ALA 2111    & D & 324.7 & 5.18 & 1.17 &$ -0.58$ \\  
Cl* NGC 6397 ALA 2111    &   & 327.3 & 3.64 & 1.20 &$ -0.58$ \\  
\\                          
NGC 6397 211             & G & 510.5 & 4.25 & 1.14 &$ -0.1$  \\  
NGC 6397 211             &   & 578.2 & 2.20 & 1.46 &$ -0.1$  \\
\\                          
NGC 6397 603             & G & 510.5 & 3.61 & 1.34 &$ -0.1$  \\   
NGC 6397 603             &   & 578.2 & 1.62 & 1.69 &$ -0.1$  \\   
\\                          
HD 218502                & D & 324.7 & 6.57 & 1.58 &$ -0.60$ \\  
HD 218502                &   & 327.3 & 4.62 & 1.45 &$ -0.53$ \\ 
\hline
\end{tabular}
\end{table*}

\begin{table}
\caption{Atmospheric parameters for the giant stars in NGC 6752
\citep{Yong} and NGC 6397 \citep{gratton82}. \label{gratton}}
\begin{tabular}{lcccc}
\hline\hline\noalign{\smallskip}
Star  & T$_{\rm eff}$ & $\log g$   & [Fe/H] & $\xi$\\
      &     K         & [cgs]      &  dex   & \kms \\
\hline\noalign{\smallskip}
Cl* NGC 6752 YGN 30 & 4943 & 2.42 & -1.62 & 1.27\\
NGC 6397 211 & 4210 & 0.80 & --2.0 & 3.0 \\  
NGC 6397 603 & 4400 & 1.50 & --2.0 & 2.6 \\ 
\hline 
\end{tabular}
\end{table}

\begin{table*}
\caption{3D corrections.\label{corr3D}}

\begin{tabular}{lrrrrrrr}
\hline\hline
\noalign{\smallskip}
model &         EW(pm)  &\multicolumn{3}{c}{3D--\mD}&\multicolumn{3}{c}{3D--\xx}\\               
 &   & $\xi=0.5$ & $\xi=1.0$ &$\xi=1.5$ & $\xi=0.5$ & $\xi=1.0$ & $\xi=1.5$\\
\hline\noalign{\smallskip}
\multicolumn{8}{c}{\ion{Cu}{i} 324.7\,nm}\\
\hline
\noalign{\smallskip}
d3t63g40m10n01&113.3&  $-0.623$&$-0.415$ &$-0.107$&$ -0.573$& $-0.392$& $-0.115$\\
d3t63g40m10n01&126.6&  $-0.485$&$-0.333$ &$-0.067$&$ -0.418$& $-0.284$& $-0.052$\\
d3t63g40m10n01&6.57 &  $-0.382$&$-0.276$ &$-0.197$& $-0.365$& $-0.256$& $-0.175$\\  
d3t63g45m10n01&113.3&  $-0.411$&$-0.237$&$ 0.021$&  $-0.405$& $-0.262$&$-0.041$\\
d3t63g45m10n01&126.6&  $-0.322$&$-0.192$&$ 0.030$&  $-0.298$& $-0.190$&$-0.005$\\
d3t63g45m10n01&6.57 &  $-0.325$&$-0.232$ &$-0.162$& $-0.362$& $-0.267$& $-0.194$\\  
d3t63g40m20n01&113.3&  $-0.929$&$  -0.652$&$  -0.323$&$  -1.228$&$  -0.995$&$  -0.701$\\
d3t63g40m20n01&126.6&  $-0.790$&$  -0.517$&$  -0.173$&$  -1.061$&$  -0.828$&$  -0.524$\\
d3t63g40m20n01&6.57 &  $-0.565$&$-0.468$ &$-0.400$& $-0.768$& $-0.649$& $-0.560$\\  
d3t63g40m20n01&5.34 &  $-0.525$&$-0.464$ &$-0.420$& $-0.676$& $-0.607$& $-0.555$\\   
d3t63g40m20n01&4.40 &  $-0.505$&$-0.462$ &$-0.430$& $-0.635$& $-0.589$& $-0.553$\\  
d3t63g45m20n01&113.3&  $-0.599$&$  -0.368$&$  -0.058$&$  -1.012$&$  -0.845$&$  -0.619$\\
d3t63g45m20n01&126.6&  $-0.545$&$  -0.333$&$  -0.032$&$  -0.924$&$  -0.763$&$  -0.534$\\
d3t63g45m20n01&6.57 &  $-0.350$&$-0.252$ &$-0.187$& $-0.778$& $-0.677$& $-0.599$\\  
d3t63g45m20n01&5.34 &  $-0.322$&$-0.259$ &$-0.216$& $-0.710$& $-0.649$& $-0.602$\\    
d3t63g45m20n01&4.40 &  $-0.303$&$-0.258$ &$-0.226$& $-0.674$& $-0.632$& $-0.599$\\  

\hline\noalign{\smallskip}
\multicolumn{8}{c}{\ion{Cu}{i} 327.3\,nm}\\
\hline
\noalign{\smallskip}
d3t63g40m10n01&113.1&$-0.677$ &$  -0.423$ &$  -0.116$ &$  -0.654$ &$  -0.421$ &$  -0.132$\\
d3t63g40m10n01&104.4&$-0.699$ &$  -0.443$ &$  -0.158$ &$  -0.687$ &$  -0.451$ &$  -0.173$\\
d3t63g40m10n01&4.62 &$-0.306$ &$-0.256$ &$-0.219$ &$-0.281$ &$-0.231$ &$-0.194$\\
d3t63g45m10n01&113.1&$-0.446$&$-0.243$&$ 0.016$&$-0.459$&$-0.287$&$-0.057$\\ 
d3t63g45m10n01&104.4&$-0.476$&$-0.264$&$-0.024$&$-0.507$&$-0.326$&$-0.098$\\ 
d3t63g45m10n01&4.62 &$-0.289$ &$-0.243$ &$-0.209$ &$-0.314$ &$-0.268$ &$-0.233$\\
d3t63g40m20n01&113.1&$-0.831$&$  -0.558$&$  -0.241$&$  -1.139$&$  -0.928$&$  -0.648$\\
d3t63g40m20n01&104.4&$-0.867$&$  -0.589$&$  -0.326$&$  -1.202$&$  -0.984$&$  -0.704$\\
d3t63g40m20n01&4.62 &$-0.492$ &$-0.444$ &$-0.409$ &$-0.625$ &$-0.573$ &$-0.534$\\
d3t63g40m20n01&3.94 &$-0.496$ &$-0.458$ &$-0.431$ &$-0.619$ &$-0.580$ &$-0.550$\\
d3t63g40m20n01&2.60 &$-0.575$ &$-0.553$ &$-0.537$ &$-0.686$ &$-0.665$ &$-0.648$\\
d3t63g45m20n01&113.1&$-0.621$&$  -0.381$&$  -0.080$&$  -0.965$&$  -0.821$&$  -0.619$\\
d3t63g45m20n01&104.4&$-0.599$&$  -0.340$&$  -0.087$&$  -1.000$&$  -0.850$&$  -0.655$\\
d3t63g45m20n01&4.62 &$-0.300$ &$-0.250$ &$-0.215$ &$-0.673$ &$-0.626$ &$-0.591$\\
d3t63g45m20n01&3.94 &$-0.291$ &$-0.251$ &$-0.224$ &$-0.656$ &$-0.620$ &$-0.592$\\ 
d3t63g45m20n01&2.60 &$-0.274$ &$-0.251$ &$-0.235$ &$-0.630$ &$-0.610$ &$-0.594$\\
\noalign{\smallskip}
\hline
\end{tabular}
\end{table*}

\end{document}